# A NEW THREE-STAGE CURRICULUM LEARNING APPROACH TO DEEP NETWORK BASED LIVER TUMOR SEGMENTATION

*Huiyu Li[1], Xiabi Liu[1,]\*, Said Boumaraf[1], Weihua Liu[1] , Xiaopeng Gong[1], Xiaohong Ma[2]*

[1]Beijing Lab of Intelligent Information Technology, School of Computer Science, Beijing Institute of Technology, Beijing, 100081, China
[2]National Cancer Center/National Clinical Research Center for Caner/Cancer Hospital, Chinese Academy of Medical Sciences and Peking Union Medical College, Beijing, 100021, China

## ABSTRACT

Automatic segmentation of liver tumors in medical images is crucial for the computer-aided diagnosis and therapy. It is a challenging task, since the tumors are notoriously small against the background voxels. This paper proposes a new three-stage curriculum learning approach for training deep networks to tackle this small object segmentation problem. The learning in the first stage is performed on the whole input to obtain an initial deep network for tumor segmentation. Then the second stage of learning focuses the strengthening of tumor specific features by continuing training the network on the tumor patches. Finally, we retrain the network on the whole input in the third stage, in order that the tumor specific features and the global context can be integrated ideally under the segmentation objective. Benefitting from the proposed learning approach, we only need to employ one single network to segment the tumors directly. We evaluated our approach on the 2017 MICCAI Liver Tumor Segmentation challenge dataset. In the experiments, our approach exhibits significant improvement compared with the commonly used cascaded counterpart.

## 1. INTRODUCTION

Liver cancer is the second most common cause of cancer death worldwide. In clinical practices, segmenting malignant tissues is a prerequisite step for final cancer diagnosis and treatment planning. However, manual segmentation is time-consuming and poorly reproducible. An accurate and automatic method of liver tumor segmentation is highly desirable.

In recent years, deep learning has dramatically improved the performance of liver and tumor segmentation [1]. The best performing methods are usually based on U-net [2] architecture with residual connections [3]. For the training of U-net framework, the cascade approach is commonly applied to firstly segment the liver and then segment the tumor inside the liver. Christ et al. [4] applied two cascaded U-net models for liver and tumor segmentation, respectively. The segmentation output was further refined based on 3D Conditional Random Field (3D-CRF). Chlebus et al. [5] employed two cascaded models for tumor segmentation, which are followed by an object-based post-processing step. In Bellver et al. [6], the first network focuses on the liver regions, then an independent detector localizes the tumors in the liver, and finally the tumor segmentation is performed based on the localizations. They also used 3D-CRF for post-processing. Han [7], the winner of the first round of 2017 MICCAI Liver Tumor Segmentation (LiTS) challenge, developed two cascaded networks working in 2.5D for the liver and tumor segmentation, respectively. Li et al. [8] proposed a novel hybrid densely connected U-net called H-DenseUNet. A 2D DenseUNet is used to efficiently extract intra-slice features, then a 3D counterpart is intended to hierarchically aggregate volumetric contexts. Jiang et al. [9] proposed a cascaded model composed of three networks: the liver localization network, liver segmentation network, and tumor segmentation network.

Despite their success, the above-mentioned cascaded models are still struggling in small tumor segmentation. The size of tumors is much smaller than that of whole CT volumes as well as that of liver regions. To segment the tumor from the whole CT volume or the liver region is like finding a needle-in-a-haystack. The resultant outcome is that even if the existing algorithms can segment the liver very well, the accuracy of tumor segmentation is still unsatisfied. Adding the post-processing [4] or an additional detector [6] to the two-cascade models [7], or extending the two-cascade model to three-cascade one [9], cannot effectively improve the accuracy of liver tumor segmentation. This reflects that the cascade approach could miss the specific features of tumors because of the ignorance of notoriously small size of tumors compared with the whole background as well as the liver. Training the network to better capture the essence of tumor specific features may help to handle this small object segmentation problem.

In this paper, we propose a new three-stage curriculum learning approach to tackle the problem of small object segmentation like liver tumor segmentation. The proposed approach trains the network on two scales: tumor patches and whole input volumes. The training on tumor patches help the model better capture specific features of tumors, while the training on the whole input makes the tumor specific features play role under the global context to segment the tumor from the whole input. We integrate the two scales of training in a three-stage schema, which starts training on the whole input; then continues training on the tumor patches; and finally retrains on the whole input. In this way we can mitigate the dilemma that the interested small object is drowned in a huge background. Fig. 1 illustrates our approach, which will be explained in the next section.

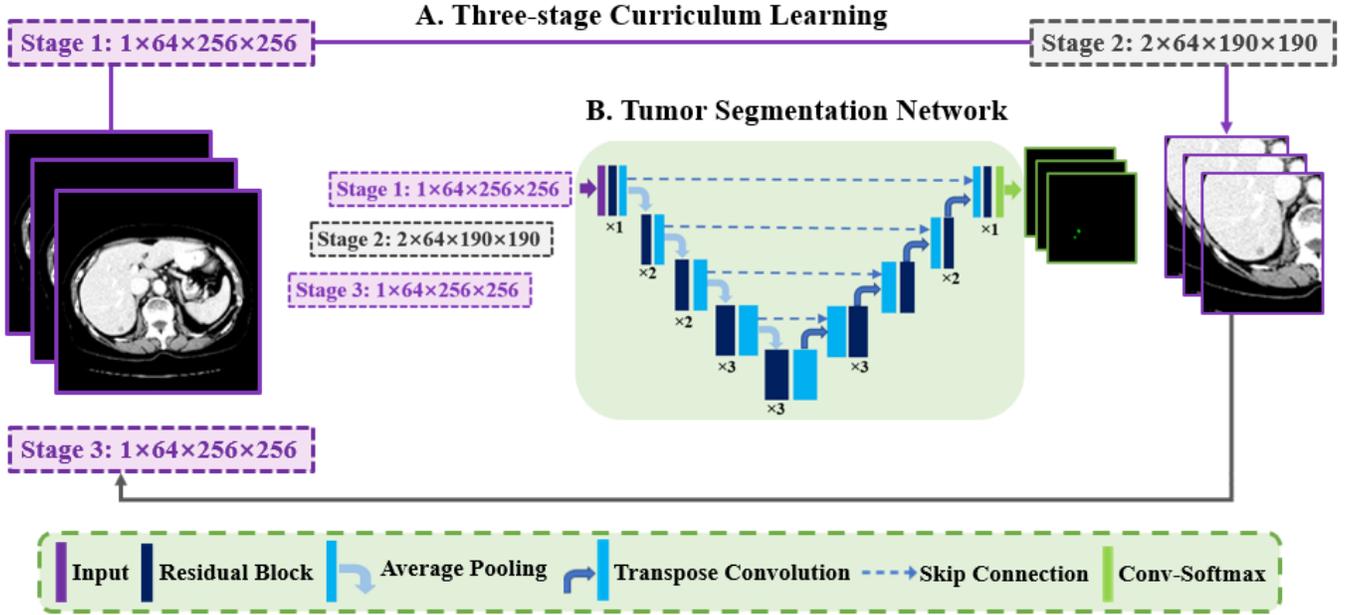

**Fig. 1**: Schematic representation of the three-stage curriculum learning approach for training the U-net based tumor segmentation network, where "× X" denotes that a residual block is repeated X times.

Our contributions are summarized as follows. 1) A new three-stage curriculum learning approach is presented to tackle the problem of small object segmentation. 2) To the best of our knowledge, this is the first introduction of curriculum learning strategy in liver tumor segmentation. 3) As far as we know, this is also the first work to segment the tumor directly, i.e. without the need of prerequisite liver segmentation.

## 2. THE PROPOSED APPROACH

The proposed approach to tumor segmentation is illustrated in Fig. 1, which is composed of a U-net based deep network and our three-stage curriculum learning approach for training this network. The details of the two components are given as follows.

### 2.1 Tumor Segmentation Network

Recently, successful works for 3D medical image segmentation are mainly based on plain U-net architecture with residual connections [10], so we also apply this architecture in this work. However, we only employ one single network to segment the tumor directly, instead of segmenting the tumor from the liver as in commonly used cascade approaches.

The employed tumor segmentation network consists of an encoder and a decoder symmetrically, which uses residual blocks as opposed to a simple sequence of convolutions. As the network goes deeper, the information is coarser and the learned features are more related to semantics. To take full advantage of the semantic features and mitigate the information loss, we elaborate the number of residual blocks for different levels of the feature maps which can be seen in Fig. 1. The network starts with some number of feature maps at the highest resolution. This number is doubled with each down-sampling operation (up to a maximum of 128) in the encoder and halved with each transposed convolution in the decoder.

### 2.2 Three-stage Curriculum Learning

Since only a very small fraction of voxels belong to a tumor in both the whole input volume and the liver region, the commonly used cascade approaches seem finding a needle-in-a-haystack and could lose some specific features of tumors. Moreover, the undesirable liver segmentation result hinders the performance of subsequent tumor segmentation.

Based on the thought above, we propose the three-stage curriculum learning approach to explore tumor related features more sufficiently and use it to segment the tumor directly on the CT volumes. As shown in Fig. 1, these three stages of learning are performed sequentially. The process and the role of each stage are described as follows.

• **Stage 1:** We start to learn the network for segmenting the tumor from the whole 3D volumes. Starting in this way, instead of firstly learning on the tumor patches, can guide the learning process under the ultimate segmentation objective and afford more robustness to the final learning result.

• **Stage 2:** In this stage, we refine the model on 3D tumor patches extracted from the whole input volumes. The size of tumor patch is decided to be the maximum one for all the tumors, in order to enclose any tumor used in the training. Except the positive samples of tumor regions, we also cropped the negative samples (i.e. do not contain any tumor) with the same size. Since tumor regions become the major part of such type of training samples, the tumor specific features can be learned sufficiently and strengthened in the segmentation network.

• **Stage 3:** The training process of this stage is same as that of the first stage, but starting from the model outputted from Stage 2. This stage aims at transferring the knowledge learned from the tumor patches to the whole input volumes, in order that the tumor specific features and the global context can be integrated ideally under our final purpose of segmenting the tumor from the whole input.

## 3. EXPERIMENTS

### 3.1 Experimental Setup

**Dataset and Preprocessing.** We use the 2017 MICCAI Liver Tumor Segmentation Challenge (LiTS) dataset to train and evaluate the proposed approach (https://competitions.codalab.org/competitions/17094#participate). The training data contains 130 CT scans, which have the same resolution of 512×512 pixels but with varied slice thicknesses. The ground truth is provided for the training set. During the experiments, we divide the scans randomly into 2 non-overlapping subsets, 90% for training and 10% for testing, respectively.

In medical image segmentation, data preprocessing is a prerequisite step for effective network training. We make use of spacing interpolation, window transform, effective range extraction and sub-image generation. For sub-image generation, we apply a patch size of 64×256×256 to optimize the trade-off between the available GPU memory and the contextual information retained in the input patches. We finally get 11541 of whole input volumes, 40957 positive samples and 21189 negative samples of tumor patches after preprocessing. Our source code of these preprocessing steps is provided at https://github.com/Huiyu-Li/Preprocess-of-CT-data.

**Parameter Settings.** The tumor segmentation network is implemented with the PyTorch framework. The network was trained with the batch size of 1 and 2 for whole inputs and tumor patches, respectively. The learning rate was set to $10^{-3}$ and reduced by a factor of 10 for each subsequent stage of curriculum learning.

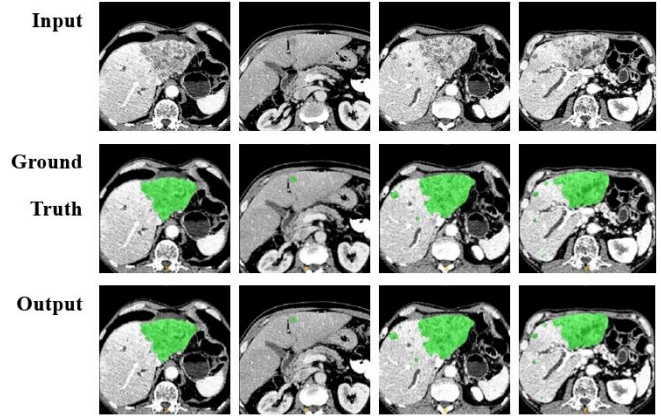

**Fig. 2**: Examples of automatic liver tumor segmentation results from the proposed three-stage curriculum learning approach, where each row from top to bottom represents input image, ground truth, and segmentation result, respectively, and the tumors are indicted in green.

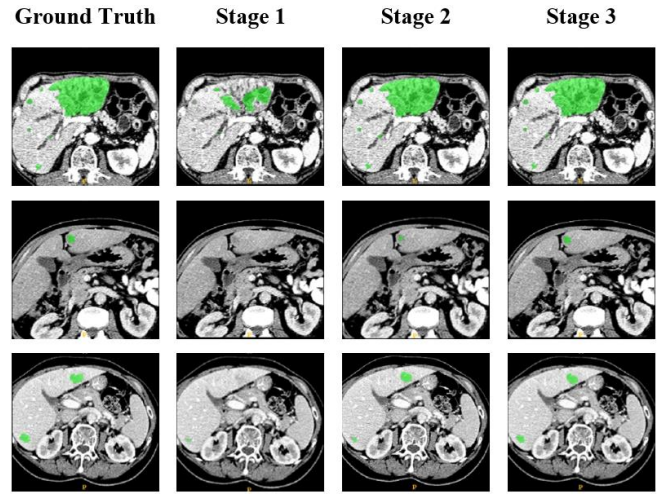

**Fig. 3**: Examples of liver tumor segmentation results from three stages of our curriculum learning approach, where each column from left to right represents ground truth, segmentation results of three stages learning, respectively, and the tumors are indicted in green.

**Evaluation Criteria.** We evaluate the performance of the proposed approach using Dice Score (DS), which consists of Dice per Case (DC) and Dice Global (DG), Volumetric Overlap Error (VOE), Relative Volume Difference (RVD), Average Symmetric Surface Distance (ASSD), Maximum Surface Distance (MSD), and Root Means Square symmetric surface Distance (RMSD) [1]. A perfect segmentation yields 1 in DC and DG, while 0 in each of other metrics (VOE, RVD, ASSD, MSD and RMSD).

**Table 1**: The performance comparisons between our approach and its counterparts on LiTS dataset.

| Approach | DC | DG | VOE | RVD | ASSD | MSD | RMSD |
|---|---|---|---|---|---|---|---|
| Three-stage Curriculum Learning | **0.822** | **0.955** | **0.235** | **0.237** | **2.458** | 41.100 | **5.149** |
| Whole-to-patch Curriculum Learning | 0.799 | 0.947 | 0.265 | 0.329 | 2.533 | 47.112 | 5.904 |
| Cascaded Architecture | 0.702 | 0.820 | 0.378 | 0.388 | 7.151 | **36.055** | 9.678 |
| Patch-to-whole Curriculum Learning | 0.633 | 0.852 | 0.457 | 0.473 | 7.513 | 46.936 | 11.249 |
| Naïve Learning | 0.671 | 0.809 | 0.421 | 0.860 | 5.115 | 37.407 | 7.965 |

### 3.2 Experimental Results

The qualitative results from our approach are visualized in Fig. 2. It shows various situations of segmentation. For the situations shown in the first two columns of the figure, there are only one tumor in each input slice but the sizes of tumors could be varied greatly across the slices. For the last two columns, there are multiple tumors in each input slice with various sizes and shapes. These cases demonstrate that the proposed approach can handle various situations and lead to satisfactory results.

For more objective and quantitative evaluation, the DS criterions of our three-stage curriculum learning are counted on all the test images and compared with those from other related learning schemas under the same tumor segmentation network as well as the same experimental setup. These compared schemas includes:

1) **Naïve Learning.** Only first stage training of our approach is considered in this schema, i.e. we naively train the tumor segmentation network on the whole input.

2) **Whole-to-patch Curriculum Learning.** In this schema, we train the network using first two stages of our approach, i.e. starting training on the whole input and end training on tumor patches.

3) **Patch-to-whole Curriculum Learning.** This schema is same as the last two stages of our approach. We start training on tumor patches and end training on the whole input.

4) **Cascaded Architecture.** In the cascaded schema, the first network is used to segment the liver and the second network is used to segment the tumor in liver. As described in Section 1, this is the commonly used approach in current liver segmentation community.

The comparison results of the aforementioned schemas are provided in Table 1 and the three stages segmentation examples of the proposed curriculum learning approach are shown in Fig. 3. From the results, we can conclude that: 1) Our proposed three-stage curriculum learning approach exhibits its promising effectiveness for the segmentation of liver tumors. It achieves the best performance for all the criterions except MSD. Especially for DC and DG, two main indicators of segmentation accuracy, we reach 0.822 and 0.955, respectively, which are obviously better than other schemas. 2) The naive learning approach achieves poor performance, which demonstrates that learning the tumor related features from the whole input volume is insignificant. 3) The whole-to-patch curriculum learning exceeds the performance of the remaining three approaches. It reflects that the first two stages of our approach (whole-to-patch curriculum learning) induce more beneficial influence than the last two stages (patch-to-whole curriculum learning). It should be noted that the last two stages of our approach have some similarities with the approach of Haarburger et al. [11], which is used to classify breast malignancy from MRI images. However, as the results shown in Table 1, such two stages of the patch-to-whole curriculum learning is not suitable to our liver tumor segmentation problem and lead to an insignificant performance. 4) Given the inferior performance of whole-to-patch curriculum learning compared with three-stage curriculum learning, the third learning stage is of paramount importance to improve the overall performance. Fig. 3 also verifies that three successive stages of the proposed curriculum learning approach are indispensable.

Benefitting from the proposed three-stage curriculum learning approach, we can use one single network to segment the tumor directly from the whole input volume. Compared with the commonly used cascaded architecture, we can obtain obvious performance improvement as shown in Table 1. Furthermore, although such curriculum learning increases the training time, it can substantially save time for tumor segmentation during testing. In our experiments, segmenting one input volume with the size of 64×256×256 can be performed under 0.75 seconds by using the single network from our approach, while 2.64 seconds are needed for the two-cascade counterpart.

### 4.CONCLUSIONS

In this paper, we have presented a new three-stage curriculum learning approach to handle small object segmentation and applied it to liver tumor segmentation. Compared with the commonly used cascaded model, the proposed approach can lead to more precise segmentation and the saving of computational time during testing. On the preprocessed dataset of LiTS, we are able to segment the liver tumor in 0.75 seconds and achieve the accuracy of 0.822 measured in DC. For the future work, we will validate the proposed approach under the rule of LiTS challenge and other popular medical image segmentation benchmarks, such as the kidney tumor segmentation challenge (KiTS) dataset.

# 5. REFERENCES

[1] Bilic P, Christ P F, Vorontsov E, et al. The liver tumor segmentation benchmark (lits)[J]. arXiv preprint arXiv:1901.04056, 2019.

[2] Ronneberger O, Fischer P, Brox T. U-net: Convolutional networks for biomedical image segmentation[C] // International Conference on Medical image computing and computer-assisted intervention. Springer, Cham, 2015: 234-241.

[3] He K, Zhang X, Ren S, et al. Deep residual learning for image recognition[C]//Proceedings of the IEEE conference on computer vision and pattern recognition. 2016: 770-778.

[4] Christ P F, Elshaer M E A, Ettlinger F, et al. Automatic liver and lesion segmentation in CT using cascaded fully convolutional neural networks and 3D conditional random fields[C]// International Conference on Medical Image Computing and Computer-Assisted Intervention. Springer, Cham, 2016: 415-423.

[5] Chlebus G, Schenk A, Moltz J H, et al. Automatic liver tumor segmentation in CT with fully convolutional neural networks and object-based postprocessing[J]. Scientific reports, 2018, 8(1): 15497.

[6] Bellver M, Maninis K K, Pont Tuset J, et al. Detection-aided liver lesion segmentation using deep learning[C]//Advances in Neural Information Processing Systems 30 (NIPS 2017): NIPS Proceedings. 2017: 1-5.

[7] Han X. Automatic liver lesion segmentation using a deep convolutional neural network method[J]. arXiv preprint arXiv:1704.07239, 2017.

[8] Li X, Chen H, Qi X, et al. H-DenseUNet: hybrid densely connected UNet for liver and tumor segmentation from CT volumes[J]. IEEE transactions on medical imaging, 2018, 37(12): 2663-2674.

[9] Jiang H, Shi T, Bai Z, et al. AHCNet: An Application of Attention Mechanism and Hybrid Connection for Liver Tumor Segmentation in CT Volumes[J]. IEEE Access, 2019, 7: 24898-24909.

[10] Yang G, Gu J, Chen Y, et al. Automatic kidney segmentation in CT images based on multi-atlas image registration[C]//2014 36th Annual International Conference of the IEEE Engineering in Medicine and Biology Society. IEEE, 2014: 5538-5541.

[11] Haarburger C, Baumgartner M, Truhn D, et al. Multi Scale Curriculum CNN for Context-Aware Breast MRI Malignancy Classification[J]. arXiv preprint arXiv:1906.06058, 2019.